# High pressure gas vessels for neutron scattering experiments.

## R. Done, O. Kirichek, B. E. Evans, Z. A. Bowden

### ISIS Facility, STFC, Rutherford Appleton Laboratory, Harwell, Didcot, UK

*The combination of high pressure techniques with neutron scattering proves to be a powerful tool for studying the phase transitions and physical properties of solids in terms of inter-atomic distances. In our report we are going to review a high pressure technique based on a gas medium compression. This technique covers the pressure range up to ~0.7GPa (in special cases 1.4GPa) and typically uses compressed helium gas as the pressure medium. We are going to look briefly at scientific areas where high pressure gas vessels are intensively used in neutron scattering experiments. After that we are going to describe the current situation in high pressure gas technology; specifically looking at materials of construction, designs of seals and pressure vessels and the equipment used for generating high pressure gas.*

## 1. Introduction

The combination of high pressure techniques with neutron scattering proves to be a powerful tool for studying the phase transitions and physical properties of solids in terms of inter-atomic distances.

Since it is defined as the ratio of force by unit area, the very high pressures can be achieved by using very small sample dimensions. However, neutron scattering studies quite often require fairly large sample volumes, and as a consequence, high pressure can only be obtained by applying large forces which implies massive mechanical devices.

Applied stress can be isotropic (hydrostatic pressure) or anisotropic (e.g. uniaxial stress). When subjected to uniaxial stress a solid is simply squeezed between two pistons. Therefore, the upper limit of stress depends essentially on the mechanical properties of the sample itself. For example, if covalent compounds are held at pressures above 30 kbar the sample may fracture. On the other hand, hydrostatic pressure does not destroy the materials (except in the case of single crystals undergoing a first order phase transition) and so the pressure limitations are simply those of present technology. In essence, pressure is applied to a substance by immersing it into a pressure transmitting medium, which is pressurised either by volume reduction (fluid and solid media) or by gas medium density increase at constant volume. The pressure medium is contained in a vessel whose wall thickness is a function of both the mechanical strength of the material and the internal dimensions. This means the vessel is critically dependent on the selected sample volume. As a result, the main difficulty encountered in neutron scattering experiments is the issue of the neutron beam having to travel through rather thick sections of the pressure vessels which contain the sample under investigation. This gives rise to unwanted absorption and coherent or incoherent background scattering. These can be reduced to workable levels by the right choice of vessel material and the optimisation of a high pressure vessel geometry design.

Alternatively, for relatively small samples sapphire or diamond anvil cells have proved to be quite effective, provided flux optimisation and background reduction are properly addressed [1]. The scope of such techniques is more limited (in practice, mostly neutron single-crystal and powder diffraction) but, they provide tremendous gain in performance up to 50 GPa (500 kbar).

In our paper we are going to review a high pressure technique based on a gas medium compression. This technique covers the pressure range up to ~0.7GPa (in special cases 1.4GPa) and typically uses compressed helium gas as the pressure medium [2]. We are going to look briefly at scientific areas where high pressure gas vessels are intensively used in neutron scattering experiments. After that we are going to describe the current situation in high pressure gas technology; specifically looking at materials of construction, designs of seals and pressure vessels and the equipment used for generating high pressure gas.

## 2. The gas pressure vessel technique development

The first practical use of a high pressure gas vessel was by Bridgman in 1932 for the measurements of metal resistivity as a function of pressure [3]. In his experiments, Bridgman used helium as a pressure-transmitting medium and reached pressures as high as 0.75 GPa at the temperature of liquid helium (4K). Later, both Wolfe and Reiffel described a gas pressure vessel based on the same principle but modified for neutron diffraction studies at pressures up to 0.56 GPa [4].

In inelastic neutron scattering the problem of intensity and background is much more serious than in diffraction measurements of structure analysis. Hence, pressure vessels with thin walls and larger sample volumes are required. Such a vessel, made of several close fitting aluminium alloy cylinders shrunk onto one another, was developed by Lechner [5]. This method allows significant improvements to the neutron efficiency of high pressure gas vessels, i.e. pressure rating to wall thickness. However this advantage has been limited by the difficulties required in the close machining and manufacturing tolerances of compound cylinders of this nature. A similar effect has been achieved by autofrettaging, a mechanical overstraining process described by Blaschko and Ernst [6]. The process generates extreme tension in the outer "skin" of the vessel and simultaneously forces the inner layers into compression. This approach has been used in the IPNS vessel design [7].

Another important direction of high pressure gas vessel technology development was the optimisation of the vessel wall material. This should have high mechanical strength as well as acceptable neutron absorption and scattering properties. It is also significant to bear in mind that the material properties may be considerably degraded by operating at non-ambient temperature conditions and exposure to chemically active samples. For these reasons, very few well-established materials are presently used for gas pressure vessels in neutron scattering applications: high-strength *aluminium alloys*, *Ti–Zr* and *CuBe$_2$* alloys [8], Maraging steel [9] and Inconel alloys [10]. In the case of the *Ti-Zr* null-matrix alloy, these two elements have coherent scattering lengths of opposite signs. It is possible therefore to prepare a material whose average coherent scattering length is equal to zero. Hence there will be no Bragg reflection from the pressure vessel contributing to the diffraction pattern. Still the extra diffuse scattering may be a problem for some applications.

While it is possible to generate pressure up to 1500MPa (15 kbar) using helium, the most "neutron friendly" materials aluminium alloys and Ti-Zr alloy, have only moderate ultimate tensile strength: ~500MPa (5 kbar) and ~700 MPa (7 kbar) respectively. High pressure gas vessels made of Maraging steel or beryllium-copper (Berylco 25) can sustain 800 MPa (8 kbar) without bursting [9], but using these materials worsen the situation with neutron adsorption and coherent or incoherent background scattering.

## 3. Applications of high pressure gas vessels in scientific research

The gas pressure vessels were developed for applications in several research fields. Some of these fields have been revealed quite recently, whereas others were attracting significant scientific interest for decades. Here we are going to present a few examples where high pressure gas vessel sample environments play a significant role.

### 3.1. Clathrate hydrates

Clathrate hydrates represent a large group of inclusion compounds, with the host framework made up of hydrogen-bonded $H_2O$ molecules and guest molecules trapped inside the polyhedral cages. Hydrates of natural gases prevail in deep-sea sediments and permafrost, and could play an important role in the formation of planetary nebulae [11, 12]. Most recently, a new class of clathrate hydrates containing hydrogen as the guest molecules has been discovered [13, 14]. Very few clathrate hydrates are stable at ambient conditions. Most need elevated gas pressures of the guest component for formation and stability. Therefore, experiments on hydrates can only be done at elevated gas pressures (in situ) or on samples recovered to ambient pressure and at liquid nitrogen temperatures, where gas hydrates can be kept in metastable state for extended periods.

A number of powder diffraction investigations were performed on gas hydrates on the D2B instrument at the ILL, Grenoble. The high strength aluminum pressure vessel has been used in the pressure range from 12 MPa (120 bar) up to 100 MPa (1kbar). The focus was on the pressure-dependent changes of structure and cage filling of various gas hydrates [15, 16]. More recently, a series of kinetic diffraction experiments were performed on the D20 instrument at the ILL to study the formation and decomposition kinetics of CH4- and CO2-hydrates [17, 18]. Unique information, in particular on the initial stage of the reactions (e.g. existence of intermediate phases), was obtained from these in situ studies.

In experiments at the Los Alamos National Laboratory, the crystal structure of the $D_2$ clathrate hydrate was determined as a function of temperature and pressure by neutron diffraction [19]. Synthesis of the clathrate was performed in a large volume (~2.5 ml) gas vessel under 220 MPa (2.2 kbar) of $D_2$ pressure at 200–270 K. The neutron diffraction data were mostly collected in two sets: on cooling from 200 K to 40 K at ~ 200 MPa (pressure was slowly decreasing from 210 MPa to 140 MPa) and during heating from 40 K to 200 K at ambient pressure. The ability of the clathrate to reversibly insert/de-insert substantial amounts of hydrogen suggests that clathrate hydrates are potential hydrogen storage materials.

Recently the high-resolution inelastic neutron scattering experiment on three samples of ternary tetrahydrofuran clathrate hydrates, containing either $H_2$ at different para/ortho concentrations, or HD, was performed on the TOSCA spectrometer at ISIS, UK [20]. It was demonstrated that the $H_2$ molecule rotates almost freely, while performing a translational motion in the cage, resulting in a paradigmatic example of the motion of a quantum particle in a non-harmonic three-dimensional potential well. Spectra obtained at different temperatures confirm an appreciable anharmonicity of the potential energy for the $H_2$ molecule in the cage.

### 3.2.   *Large isotropic negative thermal expansion*

Pressure-induced amorphization is a phenomenon of widespread occurrence among framework structures, and many recent theoretical and experimental investigations have been devoted to extending our understanding of the underlying mechanisms [21, 22]. One very good example is the remarkable behaviour of $ZrW_2O_8$, a cubic compound that contracts on heating, has been known about for the past 30 years [23]. This compound shows a large isotropic negative thermal expansion (NTE) over a wide range of temperatures (0–1050 K) [24]. The NTE behaviour in $ZrW_2O_8$ has been suggested to be due to the transverse vibration of an O atom in the W-O-Zr linkages [25, 26] and rigid unit modes [27]. Specific heat [28] and inelastic neutron scattering measurements [29] have shown the importance of low energy phonons in understanding NTE.

In an experiment on the IN6 spectrometer at the ILL, 15 g of $ZrW_2O_8$ sample has been compressed using argon gas in a high pressure vessel [30]. The use of argon gas as a pressure transmitting medium allowed the neutron scattering measurements at 160 K which is above the critical point of argon. The measurements in the cubic phase have been performed at ambient pressure, 30 MPa (0.3 kbar), 100 MPa (1.0 kbar), and 170 MPa (1.7 kbar). The experimental results provided direct experimental evidence of the large phonon softening as a function of pressure. These phonons include the acoustic, and librational and translational optic modes. It was also shown that this anomalous phonon behaviour can be responsible for the observed large negative thermal expansion in cubic $ZrW_2O_8$ over the full temperature range of 0–1050 K.

The ceramic material $ZrW_2O_8$ is well suited for use as a component in composites with tailored thermal expansion coefficients [31].

### 3.3.   *High-pressure water ices*

Water has a very complex phase diagram which consists of fourteen known crystalline phases [32] and also some amorphous phases [33]. Various gas pressure vessels were used to study the phase diagram of water. The transformation behaviour between different high pressure phases of ice is known to depend on the gas used. Helium, and also to some extent neon, is found to enter into ice I*h* and ice II (but not into ice III or V) leading to changes in the phase boundaries [34, 35]. Using argon gas as a pressure medium apparently does not change the stability range and structure of the known high pressure phases of ice [36].

The highly accurate pressure control inherent in a gas pressure system allowed the formation of metastable high pressure phases of ice IV and the ice XII which have been identified and studied in the pressure range 0.2–0.6 GPa (2 - 6 kbar) [37]. Both phases are metastable with respect to the less-dense ice V. This region of the water phase diagram thus provides a potential model system for experimental and theoretical studies of metastability.

In [38] the authors have applied inelastic neutron scattering to study a lattice dynamical of ice I*h*. The hydrostatic pressure up to 0.55 GPa (5.5 kbar) was applied by use of a gas pressure vessel and fluid nitrogen as a pressure medium. It was demonstrated that the anomalous features in the phonon dispersion are at the origin of the negative thermal expansion coefficient in ice below 60 K. Moreover, the results gave the first clear experimental evidence that pressure induced amorphization in ice occurs due to a lattice instability.

The neutron diffraction on the triple-axis spectrometer 1T1 at the LLB (CEA Saclay, France) and TOF powder diffractometer PEARL at ISIS  (RAL, UK) in combination with a gas pressure vessel device allowed to measure the lattice parameters of ordinary ice (D2O ice I*h*) up to 0.5 GPa (5 kbar) at 145 K under true hydrostatic pressure conditions [39]. In both cases Nitrogen was used as the pressure medium for which it was shown that under the above experimental conditions no clathrate hydrates are formed. The results have been compared with earlier studies and revealed a more precise equation of state of ice I$h$ at 145 K. The elastic anisotropy was found to be very small in accordance with the very small anisotropy of thermal expansion.

### 3.4. Iron arsenides superconductors at high pressure

In 1986, Bednorz and Müller discovered high temperature superconductivity in charge doped cuprate materials [40]. Unfortunately, and despite two decades of intensive research, scientists have so far failed to unambiguously identify the superconducting mechanism in these materials. The discovery of a second family of high $T_c$ materials, the iron arsenides, has therefore created a great deal of interest [41]. Unlike the insulating cuprate parent compounds, even the undoped iron arsenides are metallic, and superconductivity can be induced by either doping or pressure [42].

Single-crystal neutron and high-energy x-ray diffraction measurements have identified the phase lines corresponding to transitions among the ambient-pressure paramagnetic tetragonal, the antiferromagnetic orthorhombic, and the nonmagnetic collapsed tetragonal phases of $CaFe_2As_2$ [43]. Authors have found no evidence of additional structures for pressures of up to 2.5 GPa. Neutron scattering measurements employed an Al-alloy He-gas pressure vessel which provided hydrostatic pressure conditions. The highest pressures achieved in this experiment were 0.6 GPa.

There is no clear understanding of iron arsenides superconductivity mechanisms so far, and high pressure in-elastic neutron scattering experiments which usually require gas pressure vessel sample environment may play a significant role in the exploration of this phenomenon.

### 3.5. Heavy fermion superconductors at high pressure.

For more than two decades, the heavy fermion superconductor $URu_2Si_2$ has challenged researchers. The hidden order state with transition temperature T0 = 17.5K is characterised by a BCS-like specific heat anomaly [51] that is too large to be due solely to a small associated anti-ferromagnetic (AFM) moment [52], which is now believed to arise from internal strain due to sample defects. Several studies have shown the importance of experimental conditions during pressure measurements on $URu_2Si_2$. The most pronounced example is that when measured in helium, no AFM moment appeared below 0.5 GPa, but in the same sample a substantial AFM moment was detected at 0.45 GPa when using the less hydrostatic Fluorinert liquid [53]. The sensitivity of $URu_2Si_2$ to non-hydrostatic conditions is not unique in this regard. It was also demonstrated in the structurally related iron pnictide compound $CaFe_2As_2$ that non-hydrostatic conditions can result in phase coexistence [43]. Motivated by these earlier studies authors of Ref. [54] investigated the onset of AFM order in $URu_2Si_2$ using helium gas pressure vessel with a maximum working pressure of 1.0 GPa doubling the range measured in Ref. [53].They found that hydrostatic conditions are of paramount importance: the AFM transition is considerably sharper. Authors also suggested the existence of zero temperature multicritical point at Pc = 0.80 GPa. Unfortunately no technical details concerning the helium gas pressure vessel has been given in Ref. [54].

### 3.6. Metal – insulator transition in perovskites

Rare-earth ($R$) perovskites $RNiO_3$ provide a remarkable opportunity to study the relationship between structural and physical properties, since, by moving along the $4f$ rare-earth series, the evolution of several transport and magnetic properties can be nicely correlated with steric effects associated with the lanthanide contraction. The best example is given by the metal-insulator (MI) transition discovered for the compounds with $R \neq La$ [44], whose transition temperature $T_{MI}$ decreases when the size of the rare-earth ion increases.

The modification of the crystallographic and magnetic structures of $PrNiO_3$ associated with the decrease of $T_{MI}$ with external pressure 0.5 GPa (5kbar) have been investigated in neutron powder diffraction experiment on the DMC diffractometer at PSI [45]. In the experiment a $Ti$ and $Zr$ null-matrix alloy gas pressure vessel has been used.

The experimental results indicated that the crystallographic structure of $PrNiO_3$ does not react in the same way under application of internal and external pressures. On the other hand the magnetic structure observed below $T_{MI}$ at ambient pressure is preserved in the pressure range studied. PrNiO$_3$ is suggested to be magnetic (and probably insulating) below $T_{MI}$ at 0.47 GPa (4.7 kbar).

In the later experiment the evolution of the structural properties of $A_{12}xAx_8MnO_3$ was determined as a function of temperature, average $A$-site radius $\langle r_A \rangle$, and applied pressure by using high-resolution neutron powder diffraction [46]. The high-pressure neutron powder diffraction data were collected at pressures between 0 and 0.6 GPa (6 kbar), using the time-of-flight special environment powder diffractometer at the intense pulsed neutron source of Argonne National Laboratory, equipped with a helium gas vessel.

The metal-insulator transition, observed in the experiment was found to be accompanied by significant structural changes. Both the paramagnetic charge-localized phase, which exists at high temperatures and the ferromagnetic charge-ordered phase, which is found at low temperatures are characterized by large metric distortions of the $MnO_6$ octahedra. The distortions decrease abruptly at the transition into the ferromagnetic metal phase. These observations are consistent with the hypothesis that, in the insulating phases, lattice distortions of the Jahn-Teller type, in addition to spin scattering, provide a charge-localization mechanism.

Despite more than a decade of intensive research, a complete understanding of the physics underlying manganite properties at the microscopic level has not yet been achieved, mainly due to the complexity arising from the competition between double exchange and super-exchange, as well as the interplay between magnetic ordering, orbital ordering and Jahn-Teller effect [47, 48]. Hence neutron scattering in combination with gas pressure vessel sample environment offers promising method for experimental research in this area.

# 4. Gas pressure vessel technology

## 4.1. High pressure vessel designs

With an assorted range of high pressure scientific studies regularly undertaken using neutron scattering techniques, the associated containment vessel designs can be equally varied in both shape and size (figure 1). Geometric principles will be in the main driven by the specific technical requirements of the research under investigation. The unquestionable safety of using any such configuration will however always dictate the applicable design standards adopted for evaluating the limiting conditions of use. As many of these pressure vessels are employed to near maximum stress conditions, extensive use of advanced design tools such as finite element analysis (FEA) are utilised to gain a better understanding of how they will perform under the prescribed conditions.

Wherever possible, principle shapes such as cylindrical forms are used for the vessel designs. This offers relatively easy analysis of the extreme stresses encountered at maximum pressure applications. Unfortunately this is not always the case, and in many designs such as flat plate type vessels, more extensive solutions need to be employed for the analysis of these conditions. It is common for a compromise to be made between the amount of material needed in the design to fulfil stringent safety requirements and the important scientific parameter of getting neutrons into and out of the vessels.This is a very difficult balance to make and becomes a precept of the structural dimensions of the pressure vessels, the materials of construction and the neutron sensitivity of the samples being investigated.

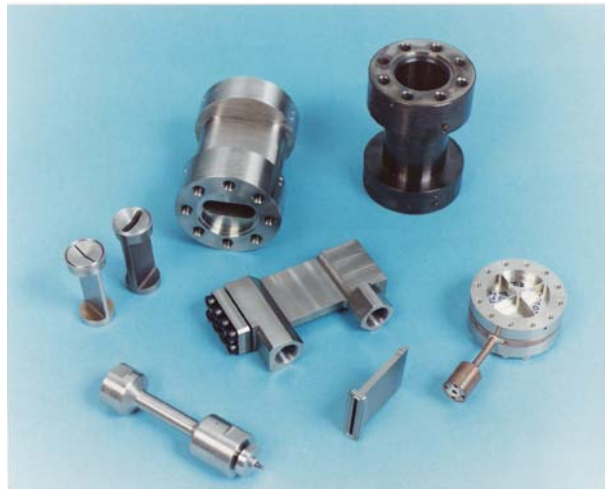

**Fig.1 A view showing a selection of high pressure vessels of differing geometries for neutron use.**

In addition to these fundamental design limits, the fabrication of the pressure vessels can itself become a challenge. By definition, the selection of high strength material alloys to withstand the enormous stresses makes processes like machining more difficult. In some cases, even joining techniques like welding, are impossible to achieve in some of these alloy grades. In addition, the scientific demands of very small internal chambers to house the experimental samples very often means that advanced techniques such as wire and spark eroding are the only methods that can be employed to create the desired geometry.

Throughout all of these processes, the designer must conform to any governing standards, ensuring that all such pressure vessels produced will be inherently safe for use. These principles dictate that not only the instantaneous conditions of use be considered, but all other critical factors need to be assessed. Examples of these features are cyclic loading, which can cause creep and fatigue effects, and even chemical erosion due to the very nature of the samples being studied. It is important that long term effects are considered at the start of any design and the appropriate limits of use defined.

Some pressure vessels are designed in such a way that reduced material thickness around a neutron entry point can be achieved (figure 2). Here the analysis of any such vessel would be dependent on advanced techniques using FEA since the complex geometries involved do not lend themselves to any standard methods. Careful control of machining standards needs to be maintained to ensure all potential stress points are reduced. A traceable set of documentation is imperative for all pressure vessels, accounting for all material data and manufacturing controls and inspection.

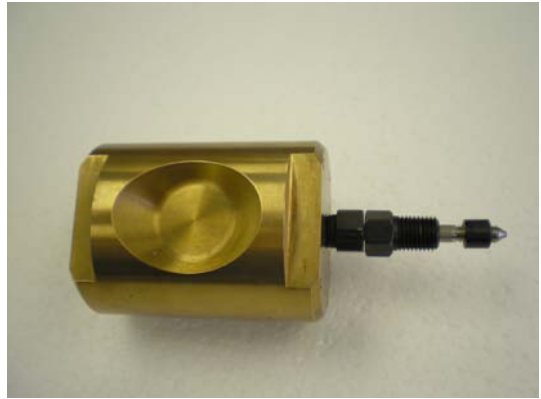

**Fig.2 A beryllium copper high pressure vessel showing a reduced area window.**

Once they are taken out of service, the eventual decommissioning of these high pressure vessels is an important factor to be considered, with detailed control over disposal based very much on the vessel's history of use. This would be the case where a pressure vessel has been used with a variety of chemical samples over a long period of time. The chemical reaction of each element can in the long term have an adverse effect on the structural material's characteristics and there may even be some cases where residual sample remains on internal surfaces. It is important therefore for companies handling the disposal of any such vessels to know the nature of the material they are supplied with in order to follow the appropriate technical and safety procedures.

Different gases can be used as the pressurising medium in the containment vessels, making sealing of any assembly difficult especially under the possible temperature conditions some are subjected to. Leakages under cryogenic conditions can cause gases to condense and freeze across the surface of the vessels making them unusable for neutrons. Leakages under high temperatures could produce high velocity, high temperature gas jets.

The use of equipment at high pressures is only achievable due to the mechanism known as autofrettaging. This is a mechanical strengthening process where pressured containers are deliberately subjected to an internal force greater than the limiting elastic strength of the material they are manufactured from. As a result, the internal material surface of any such vessel undergoes non-uniform plastic flow which gradually progresses through the containing wall as any applied pressure is further increased. When the pressure is eventually released a residual stress pattern is generated between the outer and inner surfaces, leaving the internal surfaces of the vessel in a fully compressive state and the outer in tension, but still safely below the elastic limit. The vessel can now be operated elastically up to this so called autofrettaging pressure with no further non-linear plastic flow occurring.

For monobloc cylinders, the analysis of this autofrettaging process is well known and reliable equations based on the wall thickness ratio can be used to determine safe limits of yield, autofrettage and ultimately the burst pressure (figure 3). For more complex designs, especially those with irregular geometry, a combination of these equations and non-linear finite element analysis has to be employed.

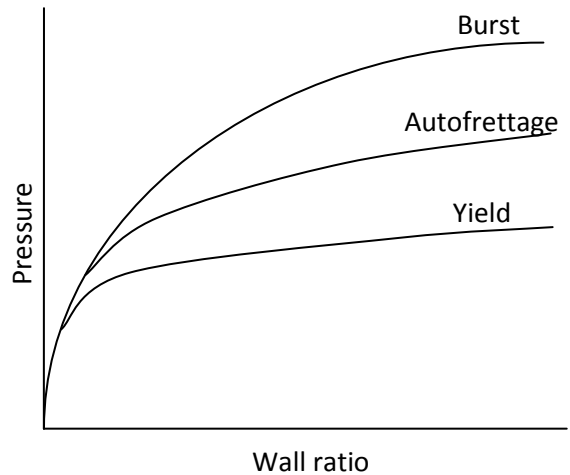

**Fig.3 Chart showing yield, autofrettage and burst pressures for monobloc cylinders.**

Finite element analysis (FEA) is a useful tool to the designer of high pressure components as it offers an insight into the actual stresses occurring within a structure subjected to the resultant forces (figure 4). In simple shapes such as cylinders and spheres, the analysis matches well with both theoretical and empirical results. Pressure vessels are rarely ideal in shape though and even the slightest change in section or blended radius to a surface can result in some startling stress concentration areas. Features such as thread forms are areas of concern and the designer is able to create models using the FEA to demonstrate how materials will react under real loaded conditions.

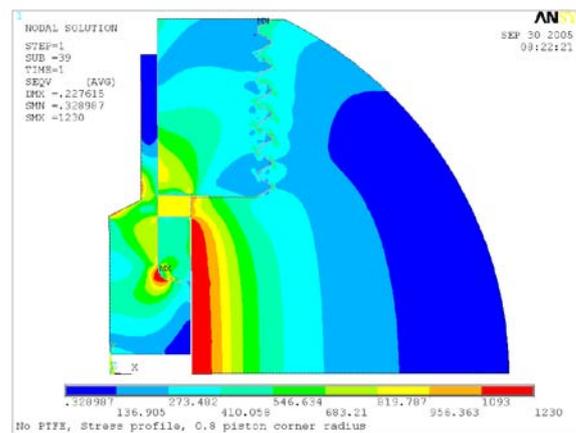

**Fig. 4 FEA stress profile of a seal arrangement and the threaded section of a pressure vessel body under compression.**

Geometric shapes are by definition three dimensional, and a vessel's feature in one plane can have an influence on other elements within the overall object. FEA is the only realistic way of making this interactive analysis of the magnitude and effect of the resultant stresses (Figure 5). Based on the deflections caused by the enormous internal pressures of the containment vessels, strains are calculated and then converted into stress for any given material characteristics. This becomes a powerful tool in recognising areas in the design of high pressure vessels where more material may be needed to prevent failure of the structure. In the case where deliberate removal of material to facilitate a neutron entry window for example, FEA will indicate the limiting pressures on that region.

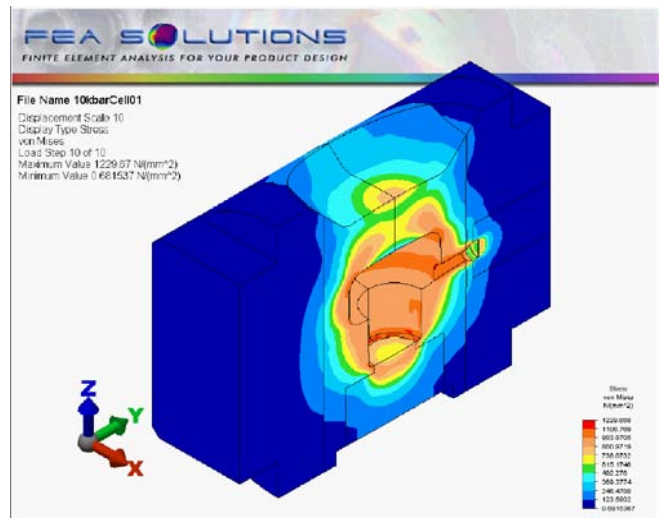

Fig.5 3D FEA stress profile of a pressure vessel body showing the highly stressed region around a window area.

Evolving designs into usable pieces of high pressure equipment involves the inevitable practicalities of feeding pressure into any such structure. Here too FEA can help by analysing the internal interactions of features such as gas entry ports where they break through into larger elements of the design (figure 6). These are areas where inevitably high stresses will exist and are especially critical in the high pressure regions.

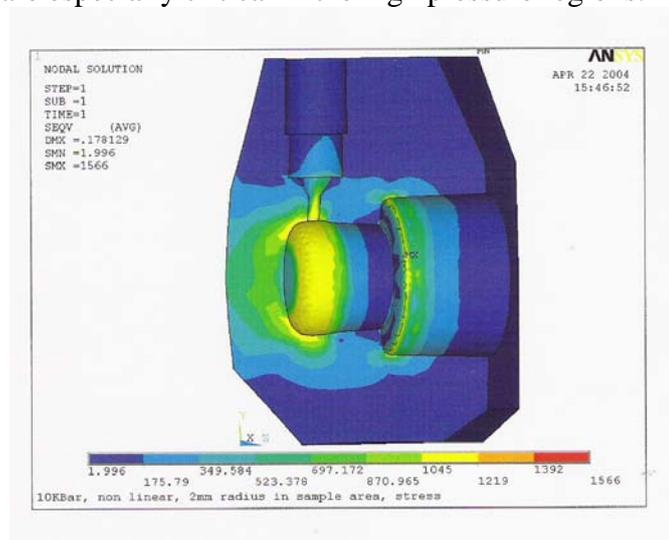

Fig.6 3D FEA of a pressure vessel showing the interaction of entry ports.

In essence, pressure vessels are designed to be compliant to the Sound Engineering Practice conditions of the Pressure Equipment Regulations 1999, working to codes such as PD5500 and ASME VIII for implementation. These define strict design limits, although the extreme pressure conditions of the equipment very often means that supporting evidence such as burst test results will be valuable as well. In addition, some empirical data, such as that relating to fatigue values, are integrated into the process to allow a more realistic working pressure rating for every vessel.

## 4.2.   Seals

The sealing of high pressure vessels present unique challenges to any design and are driven by the size and shape of the required opening into the vessel. Seal forms fall into three distinct areas:

- Gasket arrangements
- Metal to metal contact
- Unsupported area seals

### 4.2.1 Gasket arrangements

Generally, the higher the pressure inside a vessel the less likely a gasket type arrangement will be usable. This is a function of the gasket materials' endurance to the high loads encountered under pressure. Normal polymer based gaskets, such as O rings, are by their very nature too soft to withstand the huge extrusion forces apparent. This extrusion effect is a function of both the pressure and the seal size. As these are increased the use of gaskets decreases. Occasionally, backing rings can be adopted which limits the point at which extrusion takes place. Once occurring however, extrusion will result in a failure of the vessel's closure integrity. Solid metal gaskets manufactured from malleable materials such as pure aluminium or even gold can be used. There are even hollow metal O rings designed to withstand intense pressures which are coated with a soft material such as silver or lead on the outside to make a primary seal. This type of seal depends on the ingress of the gas pressure inside the assembly to generate sufficient sealing force against the surfaces of two mating parts of the pressure vessel.

### 4.2.2 Metal to metal contact

A more fundamental way of sealing high pressure equipment is by a metal to metal removable fitting. This is a proven design and relies on parts with mating cones to produce a localised sealed contact area through the deformation of one conical component against the other (Figure 7). High accuracy of parts and good surface finishes of the contact areas are essential. Loading against the pressure forces is provided by a large backing nut, although since the seal contact area is relatively small, the resultant forces are reasonably marginal. Special thick-walled tube and related fittings are available from specialist companies in this field to provide this type of seal. The only limit on this type of arrangement is size of opening, as the clamping forces will dramatically increase with any subsequent increase in aperture sizes. Openings tend to be limited to millimetres in diameter, making this type of fitting only really usable for pressure inlets.

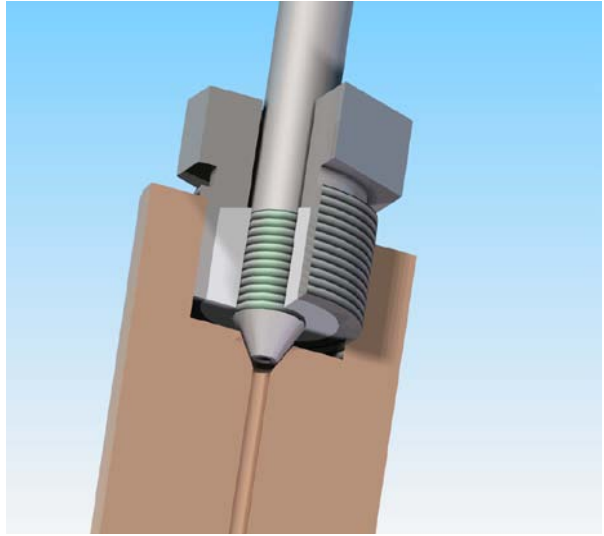

Fig.7 Standard metal to metal cone seal showing the principle of clamping and seal area.

### 4.2.3 Unsupported area seals

For openings in pressure vessels which require much larger aperture access, the obvious choice is that of an unsupported area seal. This can in principle be used up to the strength limits of the materials of construction. This type of seal can operate up to very high pressures provided there are no constraints in the essential parts of the clamping mechanism, such as the backing nut, which is there to react against the enormous seal forces involved. One version of this arrangement is also known as a Bridgman seal, named after Professor Percy Bridgman who pioneered a significant amount of the early work into high pressure research in the early 1900s [3]. Alternative designs, some of which may be more appropriate in the case of small bore diameters, are covered by Whalley et Lavergne [50].

A classic Bridgman seal requires the use of a packing type closure seal in the highly polished bore of a cylinder. It is typically described as an unsupported area seal and depends on the pressure medium producing an over-pressure in the seal parts which in turn causes deformation of these components into the gaps between the seal and bore. This prevents leakage even when gas is used as the pressuring fluid (figure 8). Initial deformation of the packing seal with some type of mechanical pre-load device makes a primary soft-seal by compressing material against both the bore of the pressure vessel and the supporting piston. This piston is sometimes referred to as the mushroom plug. The whole arrangement consists of an assembly held in place by a retaining backing nut.

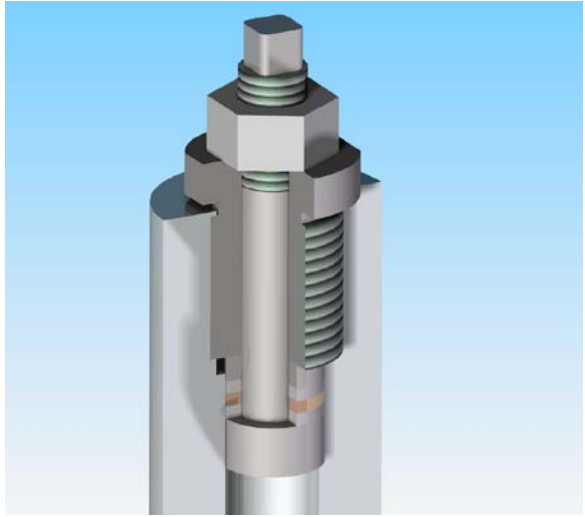

Fig.8 Typical Bridgman seal arrangement showing a three part seal assembly and the supporting backing nut.

The seal itself normally consists of three individual parts, with sacrificial soft seals either side of a slightly harder principle seal. These are fundamentally small round discs and are made from a malleable material such as PTFE or lead surrounding the principle seal, nominally manufactured from annealed copper with thin lead plating on its outside surface. After an initial closure is made with the sacrificial discs, the internal pressure can then be increased within the pressure vessel. At this point the copper disc is brought into use making the main leak free sealing state. These seals work extremely well at cryogenic temperatures [49], although selection of the most suitable materials for these applications is essential. In principle, ductile materials are essential in these assemblies as any brittleness to the seal's properties would cause immediate leakage around both the mushroom plug and the bore of the containment vessel at the intensely low temperatures involved. Under the cryogenic conditions it is vital that no leakage takes place as any released gas at those temperatures would instantly condense and freeze on any external surfaces of the pressure vessels. This could render the vessel useless for the science being undertaken and would deliver unreliable results.

### 4.3.  Materials

The selection of materials for high pressure vessel design is very much driven by their inherent properties of strength and ductility. In addition however, the caveat exists where those materials used for parts which are located in the neutron beam should also be neutron compatible, minimising the amount of absorption and scattering of neutrons. As such, many of the really high strength materials such as maraging steels, which would be able to withstand the enormous pressures, are unsuitable candidates for vessel designs due to their undesirable neutron characteristics. Generally, high strength aluminium alloys, beryllium copper, inconel and a special null matrix alloy of titanium/zirconium are the best materials of choice under these applications. Whilst these do not offer ideal conditions for neutron research, a compromise has to be made between what can be achieved structurally to withstand the stresses involved and what can be achieved scientifically with attenuated neutrons. There has also been some recent materials development work of so called 'magic super-alloys' which may show promise for the future in this field. These materials are thought to be only weakly magnetic and have mechanical properties similar to that of maraging steel. Their neutron compatibility is as yet unknown however.

The problems of optimising neutron pressure vessel designs are not limited to those parts in the neutron beam though. Other materials may be worth considering. In the case of 'pillar' or 'McWhan' type clamped designs, maraging steel for example can be used in the construction, provided windows are left accessible to the neutron beam.

Temperatures play an important role in much of the study being undertaken and as a consequence the materials selected must be able to withstand significant temperature changes. Not only will properties such as the strength change with temperature, but also factors such as elongation, brittleness and conductivity are also changeable. The use at cold temperatures offers brittleness issues, where stress cycling can be a problem. The use at elevated temperatures can often limit what materials can be used because of annealing limits and general reduction in strength.

Apart from the primary requirements of ultimate strength of the material being selected, some longer term properties also need to be considered. As the application of pressures will tend to be made in a cyclic manner, where the pressure will be increased and then later reduced to zero, fatigue characteristics become critical. This is a complex area to analyse as the mechanisms of fatigue are dependent on not only the range of pressures but also the rate by which they are changed, along with other critical factors such as the geometric features of the containment vessel. To this end, both empirical and technical test data must be used to determine the life expectancy of any vessel being used for a prolonged period. Where elevated temperatures are involved, this further complicates the issue as the creep properties of the materials being used become important. Here too it is prudent to advocate the use of test data alongside known empirical values for this feature.

It is important to recognise that as mechanical components, high pressure vessels and their associated equipment will undergo some gradual aging due to use. Since each part of any system will be operating at near maximum loading over an extended period, factors such as creep and fatigue play an important part in the correct maintenance programme. Any moving parts, such as the dynamic seals in the intensifier assemblies, will be prone to wearing characteristics and will have to be easily replaced periodically to provide reliable pressure operation. Since any such seals will generally be made from softer alloys then the surrounding structures, they tend to sacrificially wear away and need replacing more frequently. Metal to metal conical seals in particular are vulnerable to poor sealing if they are repeatedly disassembled and then reconnected. Re-machining of the contact surfaces is a regular occurrence with this type of assembly.

## 4.4.    Equipment used for generating high pressure gas

Pressure generation is made with the use of purpose built two or three stage gas intensifier pumping units. These are generally mounted inside a mobile trolley for practical reasons, allowing the equipment to be moved as close as possible to any containment vessels in use (Figure 9). A variety of gases can be applied as the pressure transmitting medium, although where possible inert gases tend to be selected.

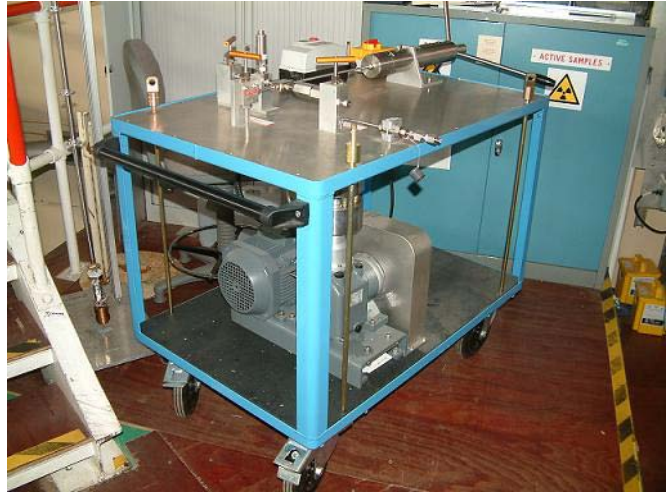

Fig.9 Typical intensifier system mounted inside its mobile trolley assembly.

The first stage of the pumping system is normally a commercially available electrically driven diaphragm pump which takes a standard gas bottle inlet supply and compresses this into the main intensifier body to an intermediate pressure (Figure 10). This is achievable through the process of cyclic deflections of a relatively thick diaphragm on one side of the pump with hydraulic oil, which repeatedly compresses the gas on the opposing side.

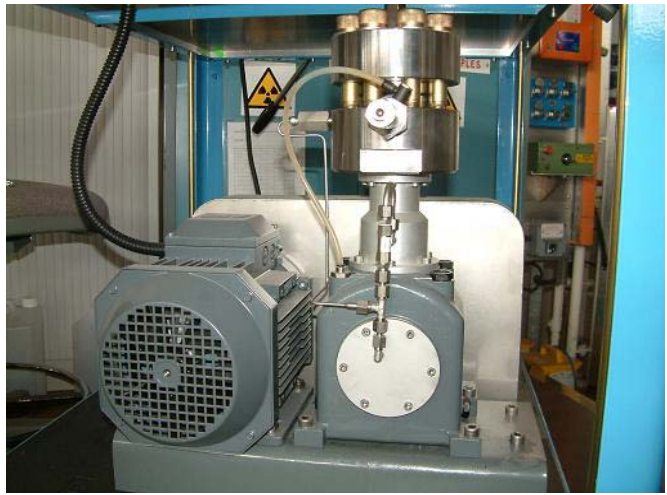

Fig.10 First-stage compressor with its hydraulic oil system at the base and the high pressure gas outlet at the top.

A second stage piston driven intensifier is then operated either manually through a capstan wheel drive system or, as some systems have, through a motorised linkage which is controlled through a feedback loop from the pressure transducer outlet. This allows compressive pressures up to the rating of the system to be generated in any containment vessel connected to the circuit. A series of valves in the circuit allow the system to be recharged with gas without any loss of pressure to these containment vessels during use (Figure 11) This is an essential requirement during low temperature applications in particular as any corresponding pressure drop due to the thermal changes will need to be recharged.

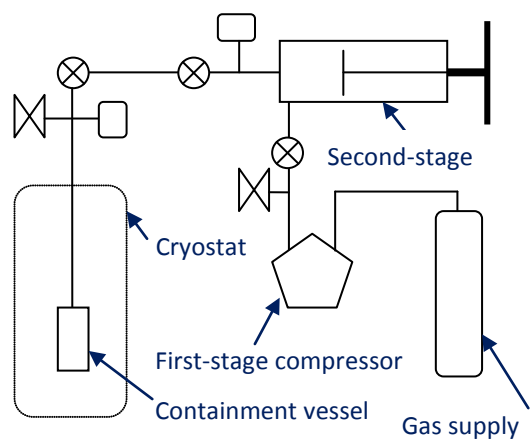

Fig.11 Typical diagram showing the complete circuit of a pressure generator system.

Appropriate pressure relief valves are fitted to prevent the accidental over pressurisation of the system which could occur within the lower pressure circuit parts during maximum pressure generation. The recorded pressure is monitored through both high pressure transducers on the intensifier side of the circuit and a similar transducer on the pressure vessel side which would also have its own shut-off valve and bursting disc protection assembly included as an integral part of its design.

### 4.5. *Pressure vessel testing, use, maintenance and associated health and safety issues*

A safety pressure test above the working conditions of any vessel is important and needs to be adopted as mandatory. It should normally be undertaken using hydraulic oil as the pressure medium, thereby reducing the stored energy contained in the assembly; which for a gas application could be enormous. To protect personnel in the event of potential failures of the equipment, all pressure testing must be carried out in a test bunker or suitably reinforced chamber. This should have remote viewing only during the test procedure, which could be done through cameras or viewing windows. All valve operations must be made from externally positioned control panels.

As recent scientific studies have developed, a greater demand is being placed on the use of high pressure vessels with hydrogen gas. This introduces a whole new series of safety considerations, not only due to the nature of handling a flammable gas at this level, but also because of the atomic effect hydrogen can have on certain materials. Hydrogen embrittlement is a typical metallurgical result and can manifest itself rapidly in the crystal structure of susceptible materials, dramatically reducing both the strength and fatigue characteristics of pressure vessels. Research into the use of surface coatings to protect high strength materials from the ingress of hydrogen is still relatively new and it will require further intense study before profound conclusions can be made about their use.

## 5. Acknowledgments

We are grateful to Jean-Michel Mignot, Burkhard Annighofer and Michael Meißner for their contribution and valuable discussions.

The project is supported by the European Commission under the 6th Framework Programme through the Key Action: Strengthening the European Research Area, Research Infrastructures. Contract n°: RII3-CT-2003-505925 ʳ

# References

[1]    Goncharenko I N and Loubeyre P 2005 *Nature* **435** 1206
[2]    Bailey I F 2003 *Z. Kristallogr.* **218** 84
[3]    Bridgman P W 1932 *Physical Review* **39** 858
[4]    Wolfe G H, Reiffel L AFOSR 1963 (*ARF-1183-4*, final report) 2641
[5]    Lechner R 1966 *Rev. Sci. Inst.* **37** 1534
[6]    Blaschko O and Ernst G 1974 *Rev. Sci. Inst.* **45**, 526
[7]    Jorgensen J D, Pei Shiyou, Lightfoot P, Hinks D G, Veal B W, Dabrowski B, Paulikas A P and Kleb R 1990 *Physica C* **171**, 93
[8]    Bloch D and Voiron J 1984 *Condensed Matter Research using Neutrons* (*NATO ASI Series* B **112**) ed. S.W. Lovesey and R. Scherm (New York: Plenum) 39
[9]    Bao W, Broholm C and Trevino S F 1995 *Rev. Sci. Inst.* **66,** 1260
[10]   Bailey I F, Performed R B, Dreyer J W and Gray E 2004 *Int. J. High Press. Res.* **24** 309
[11]   Sloan E D 1998 *Clathrate Hydrates of Natural Gases* Marcel Dekker New York
[12]   Loveday J S et al 2001 *Nature* **410** 661
[13]   Dyadin Y A et al 1999*Mendeleev Commun.* **5** 209
[14]   Mao W L, Mao H K, Goncharov A F, Struzhkin V V, Guo Q, Hu J, Shu J, Hemley R J, Somayazulu M and Zhao Yu 2002 *Science* **297** 2247
[15]   Chazallon B and Kuhs W F 2002 *J. Chem. Phys.* **117** 308
[16]   Klapproth A, Goreshnik E, Staykova D, Klein H and Kuhs W F 2003 *Can. J. Phys.* **81** 503
[17]   Staykova D K, Kuhs W F, Salamatin A N and Hansen  T 2003 *J. Phys. Chem.* B **107** 10299
[18]   Kuhs W F, Genov G, Staykova D K and Hansen T  2004 *Phys. Chem. Chem. Phys.* **6** 4917
[19]   Lokshin K A, Zhao Yu, He D, Mao W L,  Mao H K, Hemley R J, Lobanov M V and Greenblatt M 2004 *Phys. Rev. Lett.* **93** 125503
[20]   Ulivi L, Celli M, Giannasi A, Ramirez-Cuesta A J and Zoppi M 2008 *J.Phys.: Condens. Matter* **20** 104242
[21]   Kruger M B and Jeanloz R 1990 *Science* **249** 647
[22]   Tse J S, Klug D D, Ripmeester J A, Desgreniers S and Lagarec K  1994 *Nature* **369** 724
[23]   Martinek C and Hummel F A  1968 *J. Am. Ceram. Soc.* **51** 227
[24]   Perattoni C A and  da Jornade J A H 1998 *Science* **280** 886
[25]   Mary T A, Evans J S O, Vogt T and Sleight AW 1996 *Science* **272** 90
[26]   Evans J S O, Mary T A, Vogt T, Subramanian M A and Sleight AW 1996 *Chem. Mater.* **8** 2809
[27]   Pyrde A K A, Hammonds K D, Dove M T, Heine V, Gale J D and Warren M C 1996 *J. Phys. Condens. Matter* **8** 10 973
[28]   Ramirez A P, Kowach G R 1998 *Phys. Rev. Lett.* **80** 4903
[29]   Ernst G, Broholm C, Kowach G R and Ramirez A P 1998 *Nature* **396** 147
[30]   Mittal R, Chaplot S L, Schober H, and Mary T A 2001 *Phys. Rev. Lett.* **86** 4692
[31]   Holzer H and Dunand  D C 1999 *J. Mater. Res.* **14** 780
[32]   Salzmann C G, Radaelli P G, Hallbrucker A, Mayer E and Finney J L 2006 *Science* **311** 1758
[33]   Mishima O and Stanley H E 1998 *Nature* **396** 329
[34]   Londono D, Finney J L and Kuhs W F  1992 *J. Chem. Phys.* **97** 547
[35]   Lobban C, Finney J L and Kuhs W F 2002 *J. Chem. Phys.* **117** 3928
[36]   Lobban C, Finney J L and Kuhs W F 2000 *J. Phys. Chem.* **112** 7169
[37]   Lobban C, Finney J L and Kuhs W F 1998 *Nature* **391** 268
[38]   Strässle Th, Klotz S, Loveday J S and Braden M 2005 *J Phys.: Condens. Matter* **17** S3029
[39]   Strässle Th, Saitta A M, Klotz S and Braden M 2004 *Phys. Rev. Lett.* **93** 225901
[40]   Bednorz J G and Müller K A 1986 *Zeitschrift für Physik B Condensed Matter* **64** 189
[41]   Kamihara Y, Watanabe T, Hirano M and Hosono H 2008 *J. Am. Chem. Soc.* **130** 3296
[42]   Kimber S A J et al 2009 *Nature Materials* **8** 471
[43]   Goldman A I et al 2009 *Phys. Rev. B* **79** 024513
[44]   Lacorre P, Torrance J B, Pannetier J, Nazzal A I, Wang P W and Huang T C 1991 *J. Solid State Chem.* **91** 225
[45]   Medarde M, Mesot J,  Lacorre P,  Rosenkranz S,  Fischer P and Gobrech K 1995 *Phys. Rev. B*

**52** 9248

[46]  Radaelli P G, Iannone G, Marezio M, Hwang H Y, Cheong S W, Jorgensen J D and Argyriou D N 1997 *Phys Rev B* **56** 8265

[47]  Ding Y et al 2009 *Phys. Rev. Lett.* **102** 237201

[48]  Fuchigami K,  Gai Z,  Ward T Z,  Yin L F, Snijders P C, Plummer E W and Shen 2009 J *Phys. Rev. Lett.* **102** 066104

[49]  Done R, Chowdhury M A H, Goodway C M, Adams M and Kirichek O 2008 *Rev. Sci. Inst.* **79** 026107

[50]  Whalley E and Lavergne A 1976 *Rev. Sci. Inst* **47** 136

[51]  Maple M B et al 1986 *Phys. Rev. Lett.* **56** 185

[52]  Broholm C et al  1987 *Phys. Rev. Lett.* **58** 1467

[53]  Bourdarot F et al 2005 *Physica B* **359-361** 986

[54]  Butch N P et al 2010 *arXiv:* 1006.4140v1